\documentclass[%
reprint,
longbibliography,
showpacs,preprintnumbers,
 amsmath,amssymb,
 aps,
 pra,
]{revtex4-1}
\usepackage{graphicx}
\usepackage{dcolumn}
\usepackage{bm}
\usepackage{multirow}
\usepackage{hhline}
\usepackage{longtable}
\usepackage{xy}
\usepackage{amsfonts}
\usepackage{amsmath}
\usepackage{amssymb}

\begin{document}

\preprint{APS/123-QED}

\title{Heat transfer through dipolar coupling: Sympathetic cooling without contact}

\author{B. Renklioglu}
\email{brenklioglu@bilkent.edu.tr}%
\author{B. Tanatar}
\author{M. \"{O}. Oktel}%
\affiliation{Department of Physics, Bilkent University, Ankara, Turkey}




\date{\today}

\begin{abstract}
We consider two parallel layers of dipolar ultracold Fermi gases at different temperatures and calculate the heat transfer between them. The effective interactions describing screening and correlation effects between the dipoles in a single layer are modeled within the Euler-Lagrange Fermi-hypernetted chain approximation. The random-phase approximation is used for the interactions across the layers. We investigate the amount of transferred power between the layers as a function of the temperature difference. Energy transfer arises due to the long-range dipole-dipole interactions. A simple thermal model is established to investigate the feasibility of using the contactless sympathetic cooling of the ultracold polar atoms/molecules. Our calculations indicate that dipolar heat transfer is effective for typical polar molecule experiments and may be utilized as a cooling process.
\end{abstract}

\pacs{37.10.Mn, 67.85.Lm}
\maketitle


\section{Introduction}
In recent years, ultracold gases of polar atoms~\cite{Cr,Cr2,Cr3,Er,Er2,Dy,Dyp} and molecules~\cite{KRb,RbCs,NaK1,NaK2,LiCs1} with their long-range anisotropic interactions have attracted a great deal of interest for applications ranging from quantum information science~\cite{quant,quant2,quant3} to condensed matter physics~\cite{baranov, JinYe, Menoti, Pupillo, ospel, ospel2}.

Dipolar interaction is the dominant long range interaction in ultracold systems, if the constituent particles carry permanent electrical or magnetic dipoles. Both, atomic species with magnetic dipoles and, more recently, molecular gases with electrical dipoles have been realized experimentally. The presence of a long range and anisotropic interaction has a profound effect on the physics of the systems, leading to novel phases and previously unexplored regimes~\cite{Goral, Demler, Sansone, Pollet, Cooper}.

In most ultracold gases, the dipolar interaction is present together with the short range interactions arising from low angular momentum scattering~\cite{Rev_Giorgini, Rev_Lewenstein}. Usually the latter is dominant and a Feshbach resonance is needed to probe the regimes where dipolar effects are prominent ~\cite{Rev_Giorgini, Rev_Lewenstein}. For bosons, thermalization proceeds through S-wave scattering and most cooling methods rely on short range interactions. Cooling of spin polarized fermions is more challenging as they do not interact in the S-wave channel. To obtain degenerate spin polarized fermi gas either spin mixtures or mixture with another species is required during cooling. Dipolar interactions offer novel cooling methods due to their long range and anisotropic nature. Cooling schemes which rely on the anisotropic nature of dipolar interaction have been experimentally demonstrated for both bosons and fermions. Dipolar anisotropy connects different angular momentum channels resulting in coupling between translational and spin degrees of freedom which has been used for depolarization cooling~\cite{Hensler}, demagnetization cooling~\cite{Fattori}, and more recently spin distillation~\cite{Naylor} for bosons. In addition, universal dipolar scattering which relies both on the anisotropy and the long range of interaction has been used to cool a single component Fermi gas to degeneracy~\cite{Frisch}.

In this paper, we investigate the heat transfer between components of a system purely due to the long range nature of dipolar interactions. Specifically, we identify the parameter regime for which dipolar forces provide effective thermal contact between two, otherwise isolated, parts of the system. For this purpose, we use a model system which consists of two parallel layers of dipolar fermions separated by a distance $d$. The S-wave interactions between spin-polarized fermions is zero and all the energy transfer between the layers is due to dipolar interactions. We calculate the rate of energy transfer between the layers when there is a temperature difference between them. Furthermore, we also estimate the time scale to reach the thermal equilibrium between the layers. This time scale determines whether a cooling procedure applied to only one layer can effectively cool the other layer, providing sympathetic cooling without the adverse effects of contact.

To identify the relevant parameter regimes, we first investigate the length scales of the problem. The first length scale is provided by the geometry we consider, the distance between the layers $d$. The density of the fermions in each layer $n$, or equivalently the average distance between two particles inside the same layer $k_F^{-1}$ determines the inner dynamics of each layer through the Fermi energy $E_F=\hbar^2 k_F^2/2m$. A third length scale, $a_0$, measures the importance of the dipolar interaction. The interaction potential between two dipolar particles has the form of $V(\vec{r})=\left[C_{dd}(1-3cos^2\theta)\right]/(4\pi r^3)$ where $\theta$ is the angle between the intermolecular displacement $\vec{r}$ and the dipole orientation, and $C_{dd}$ is the dipolar coupling constant~\cite{Rev_Lewenstein}. The corresponding length scale $a_0$ is defined as
$a_0 = {C_{dd} m}/{4\pi \hbar^2}$.
The system is characterized by two dimensionless parameters ($\lambda$ and $\widetilde{d}$), derived from the above length scales. The coupling strength between the dipoles are governed across the layers by
$\widetilde{d} = d/a_0$,
and within a layer by
$\lambda =a_0 k_F$,
where $k_F=\sqrt{4\pi n}$ is the Fermi wave number.

We consider two parallel layers of ultracold dipolar Fermi gases without any tunneling between the layers. We describe the correlation effects between the dipoles in a single layer (intra-layer) using the fluctuation-dissipation theorem and the static structure factor $S(q)$ data obtained from the Euler-Lagrange Fermi-hypernetted-chain (FHNC) approximation~\cite{Abed}. We adapt the random-phase approximation (RPA) to account for the interactions across the layers (inter-layer). The energy transfer is calculated as a function of the temperature difference between the layers and the other parameters characterizing the system. We express our results in terms of a thermal conductivity between the layers for small temperature differences.

To gauge the effectiveness of thermal coupling between the layers, we calculate the time scale to reach equilibrium for small temperature difference. Our calculations indicate that the layers are strongly coupled when the distance between the layers $d$ is within a few dipolar length scales $a_0$. The amount of transferred power decreases rapidly as the well separation distance increases.

We study the system for a wide range of $\lambda$ and $\widetilde{d}$ at temperatures near the Fermi temperature $T_F$ and at lower temperatures close to $T=0.1 T_F$. In the low temperature regime although the heat conductivity decreases significantly, the equilibration lifetime remains unaffected as the specific heat in each layer also decreases.

Our calculations indicate that for dipolar thermal coupling to be significant the two layers in the system must be placed within a few $a_0$ of each other. This length scale is of the order of tens of nanometers for magnetic dipolar atoms. For instance, Dysprosium (Dy), the most magnetic atom in nature with a magnetic moment of $10\mu_B$, the value of the length scale is equal to $a_0=20.8$\,nm. This value is much smaller than the typical trapping features in ultracold atom experiments. However, the length scale ($a_0$) for ultracold polar molecules are of the order of $10^{-6}$\,m which is easily attainable in current experiments.

Ultracold polar molecules can be cooled by focusing the cooling effort onto a subsystem which is isolated from the rest of the cloud except for dipolar coupling. Such a sympathetic cooling mechanism would avoid contact between the actively cooled part of the system and the rest of the gas. For example, if one of the layers in our model is cooled evaporatively, the other layer will also be cooled without losing any particles. We also studied our system with a layer density difference to account for such a scenario.

The paper is organized as follows: In the next section we introduce our model in detail and describe our approach. In Section III, we present the results of our calculations in various parameter regimes. Section IV contains the discussion and relevant parameters for experiments. We conclude with a brief summary.

\section{The Model and Method}
\begin{figure}
\includegraphics[scale=0.6]{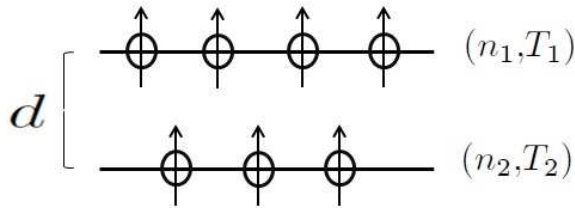}
\caption{Schematic view of the system. We consider two dipolar Fermi gas layers with different densities at different temperatures. Dipoles are oriented perpendicular to the layers which are separated by a distance $d$.}\label{model}
\end{figure}

In this study, two  parallel layers of an ultracold dipolar Fermi gas, separated by a distance $d$ is considered, as shown in Fig.~\ref{model}. The intra-layer interaction $\mathcal{V}_{11}$ within a single layer and the inter-layer interaction $\mathcal{V}_{12}$ across the layers are given by
\begin{equation}\label{V11r}
\mathcal{V}_{11}(r) = \mathcal{V}_{22}(r) = \frac{C_{dd}}{4\pi}\frac{1}{r^3}\,,
\end{equation}
and
\begin{equation}\label{V12r}
\mathcal{V}_{12}(r)=\frac{C_{dd}}{4\pi} \frac{r^2-2d^2}{(r^2+d^2)^{5/2}}\,,
\end{equation}
where the indices $1$ and $2$ denote different layers and $r$ indicates the in-plane distance between dipoles. $C_{dd}$ is the dipole-dipole coupling constant, which is $C_{dd}=\mu_0\mu^2$ for magnetic dipole moments $\mu$, and $C_{dd}=p^2/\varepsilon_0$ for electric dipole moments $p$. Here, $\mu_0$ is the vacuum permeability, $\varepsilon_0$ is the permittivity of free space. Note that $\mathcal{V}_{11}(r)$ and $\mathcal{V}_{12}(r)$ are the bare (unscreened) dipole-dipole interactions, respectively.

The Hamiltonian of the system is
\begin{equation}\label{Hamil}\begin{split}
\mathcal{H}=&-\frac{\hbar^2}{2m}\sum_i \left(\nabla_{1i}^2+\nabla_{2i}^2\right)   \\
            &+\frac{1}{2}\sum_{i,j} \Big[\mathcal{V}_{11}(|r_{1i}-r_{1j}|)+\mathcal{V}_{22}(|r_{2i}-r_{2j}|)\Big]  \\
            &+\sum_{i,j} \mathcal{V}_{12}(|r_{2i}-r_{1j}|)  \,,
\end{split}\end{equation}
where $m$ is the mass of the particles and the sums are carried out over the particles in each respective layer.

In order to describe the correlations and the resulting screened dipolar interaction within a layer, we follow Abedinpour \textit{et al.}~\cite{Abed}. The effective intra-layer interaction is obtained by using the fluctuation-dissipation theorem and static approximation as
\begin{equation}\label{V11}
V_{11}(q) =\frac{\epsilon(q)}{2n}\left[\frac{1}{S^2(q)}-\frac{1}{S_0^2(q)}\right]\,,
\end{equation}
where $\epsilon(q) = \hbar^2 q^2/2m$ is the single-particle energy. Here, $S(q)$ is the static structure factor obtained from the Euler-Lagrange Fermi-hypernetted-chain (FHNC) approximation method~\cite{Siemens,Krot}. In addition, $S_0(q)$ is the static structure factor for a non-interacting system of two-dimensional (spin-polarized) fermions.
Neglecting the correlation effects,  we use the Fourier transform of the bare
inter-layer interaction
\begin{equation}\label{V12}
V_{12}(q)=-\frac{C_{dd}}{2} q \exp(-qd)\,.
\end{equation}

\subsection{Energy Transfer Rate }

The energy transfer rate between two Fermi systems has been studied in the context of electron systems within the balance equation approach \cite{horing}, the quantum kinetic equation \cite{boiko}, and the non-equilibrium Green function method \cite{wang}. Calculations for one- and two-dimensional electron gases \cite{Tanatar, Senger} and graphene \cite{bahrami} have appeared. We adapt the energy transfer rate formulation to our double layer dipolar system characterized by layer temperatures $T_i$ and drift velocities $\upsilon_i$ and express it as
\begin{equation}\label{rate}\begin{split}
\mathcal{P}_{12}= - &\hbar\sum\limits_{q}\int\limits_{-\infty}^{\infty} \frac{d\omega}{\pi} \:  \omega  \:  |W_{12}(q,\omega,T_1,T_2)|^2 \\ \\
      &\times \left[n_B\left(\frac{\hbar \omega}{k_B T_1}\right)-n_B\left(\frac{\hbar (\omega-\omega_{12})}{k_B T_2}\right)\right]\\ \\
      &\;\mbox{Im}\;\chi_1(q,\omega ,T_1) \;\mbox{Im}\;\chi_2(q,\omega-\omega_{12} ,T_2)\,,
\end{split}\end{equation}
where $n_B(x)=1/(\exp(x)-1)$ is the Bose-Einstein distribution function and $\omega_{12}=q(\upsilon_1-\upsilon_2)$. In our calculations, the drift velocities are taken to be $\upsilon_1=\upsilon_2=0$ considering the linear regime. Here, $\mathcal{P}_{12}$ is the amount of power transferred to layer $1$ from layer $2$ per unit area.

In the above, $\chi_i(q,\omega)$ is the finite temperature two-dimensional Lindhard polarization function~\cite{Hu} for the $i$th layer. $W_{12}$ is the dynamically screened effective interaction, defined by
\begin{equation}\label{W}
W_{12}(q,\omega,T_1,T_2)=\frac{V_{12}(q)}{\epsilon(q,\omega ,T_1,T_2)} \,,
\end{equation}
in which the total dielectric function $\epsilon(q,\omega,T_1,T_2)$ is given by the random phase approximation (RPA)~\cite{Hu} as
\begin{equation}\label{epsilon}\begin{split}
\epsilon(q,\omega ,T_1,T_2)= & \left[ 1-V_{11}(q)\chi_1(q,\omega ,T_1)\right]\left[ 1-V_{22}(q)\chi_2(q,\omega ,T_2)\right] \\
                    & -\left[V_{12}(q)\right]^2\chi_1(q,\omega, T_1)\chi_2(q,\omega ,T_2)\,.
\end{split}\end{equation}
Note that our choice for $V_{11}(q)$ amounts to including intra-layer correlation effects.

The dimensionless interaction strength parameter is defined as $\lambda=k_F a_0$ where $a_0$ indicates the characteristic length scale, obtained by $a_0=C_{dd}m/(4\pi \hbar^2)$. Here, $m$ is the mass of a dipole, $k_F=\sqrt{4\pi n}$ the Fermi wave number, $n$ is the $2D$ density of a single layer.

In the sequel, we calculate the energy transfer rate under two separate conditions: (a) symmetric case where the densities of the layers are equal ($n_1=n_2$) and (b) asymmetric case where the densities are different ($n_1\neq n_2$).
In the symmetric case, dipolar gases confined to both layers of the system have the same Fermi levels ($E_{F1}=E_{F2}=E_{F}$ and $k_{F1}=k_{F2}=k_{F}$). As a result, the polarization functions, effective potential interactions and the interaction strengths of the layers become equal to each other, as $\chi_1=\chi_2$, $V_{11}=V_{22}$, $\lambda_1=\lambda_2$, respectively.

In our discussions, we use the following dimensionless quantities for the symmetric case,
\begin{align*}
Q&=\frac{q}{k_F} \,,          &  \Omega &=\frac{\hbar \omega}{E_F}\,,              &  \widetilde{d}&=d / a_0\,,\\
t&=\frac{k_B T}{E_F}\,,   &             \widetilde{\mu}&=\frac{\mu}{E_F}\,,          &  \widetilde{\chi}&=\left (\frac{\pi\hbar^2}{m}\right )\chi\,,
\end{align*}
where $E_F=\hbar^2 k_F^2/(2m)$ is the Fermi energy. Using these dimensionless quantities, the intra-layer and the inter-layer interactions can be written as
\begin{equation}\label{V11d}
V_{11}(Q)=V_{22}(Q) = \left(\frac{m}{\pi\hbar^2}\right )Q^2\left[\frac{1}{S(Q)^2}-\frac{1}{S_0(Q)^2}\right]\,,
\end{equation}
and
\begin{equation}\label{V12d}
V_{12}(Q)=- \left (\frac{m}{\pi\hbar^2}\right )2 \lambda Q \exp\left(-\lambda \widetilde{d} Q\right)\,.
\end{equation}
The dimensionless energy transfer rate $\mathrm{P}_{12}$ is
\begin{equation}\label{drate}\begin{split}
\mathrm{P}_{12} =&\ \frac{\mathcal{P}_{12}}{\left(\frac{(k_FE_F)^2}{\hbar}\right)} \\ \\
       =&-2\int\limits_{0}^{\infty}dQ\:  Q\int\limits_{-\infty}^{\infty} d\Omega \:  \Omega  \:  |W_{12}(Q,\Omega,t_1,t_2)|^2 \\ \\
      &\times \left[n_B\left(\frac{\Omega}{t_1}\right)-n_B\left(\frac{\Omega}{t_2}\right)\right]\;\mbox{Im}\;\chi_1(Q,\Omega ,t_1)\\ \\
      &\;\mbox{Im}\;\chi_2(Q,\Omega,t_2)\,.
\end{split}\end{equation}
With this scaling, unity dimensionless heat transfer $\mathrm{P}_{12}=1$ means $\left((k_F E_F)^2/\hbar \right)$ Watts of power is flowing per meter square of the system.

In the asymmetric case, the densities of the layers are different from each other ($n_1\neq n_2$), ultracold dipolar Fermi gases within the layers have distinct Fermi levels. Accordingly the interaction strengths are not equal, $\lambda_1\neq \lambda_2$. The relation between the interaction strengths and the densities of the layers is given by
\begin{equation}\label{r}
r = \sqrt{\frac{n_1}{n_2}}=\frac{\lambda_1}{\lambda_2}\,,
\end{equation}

We scale all the parameters by the Fermi energy and Fermi wave number of the first layer, and use the following relations in order to obtain the dimensionless forms of the corresponding quantities;
\begin{align*}
Q_2&=r\:   \frac{q}{k_{F1}}\,,  &  \Omega_2 &=r^2\:  \frac{\hbar \omega}{E_{F1}}\,,  \\
t_2&=r^2\:  \frac{k_B T_2}{E_{F1}}\,,   &  \widetilde{\mu_2} &=r^2\:  \frac{\mu}{E_{F1}}\,.
\end{align*}

\section{Results}

We calculate the dimensionless energy transfer rate between the layers separated by a distance $d$ for two different cases of the system, introduced in the previous section. The amount of transferred energy is obtained as a function of the temperature of one of the layers, $t_2$ for fixed $t_1$.

\begin{figure}
\includegraphics[width=8cm]{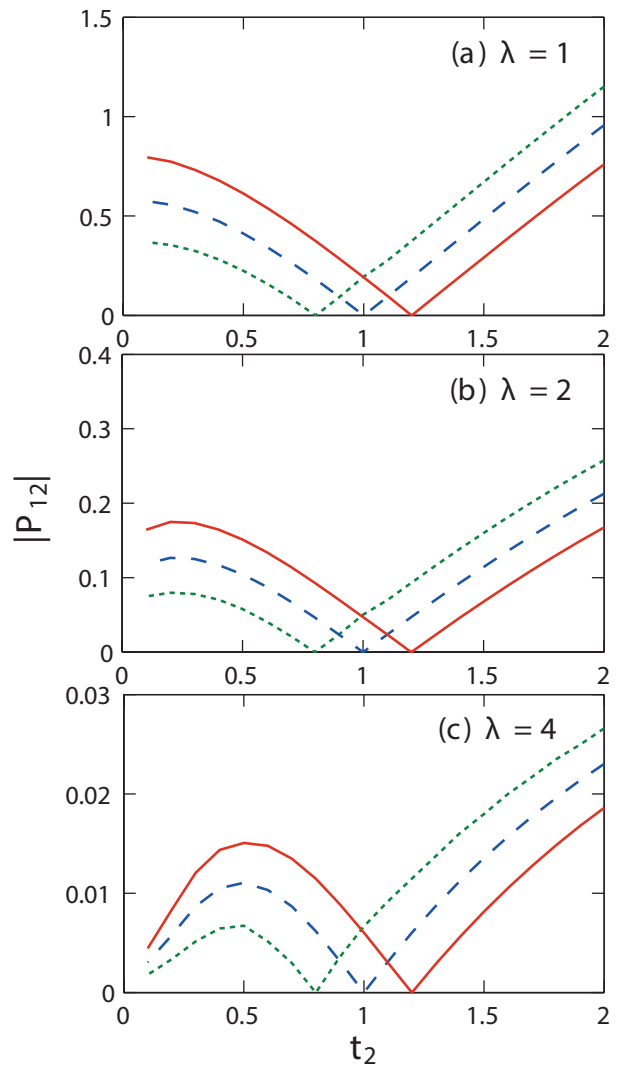}
\caption{(Color online) The absolute value of the dimensionless energy transfer rate $|\mathrm{P}_{12}|$ as a function of the temperature $t_2$ of the second layer while the other temperature $t_1$ is kept constant at three different values, indicated by solid lines (for $t_1=1.2$), dashed lines (for $t_1=1.0$), and dotted lines (for $t_1=0.8$). Here, the interaction strengths $\lambda = 1.0, 2.0, 4.0$ respectively while the dimensionless well separation distance is considered as $\widetilde{d}=1$.}\label{fig1}
\end{figure}

\subsection{Symmetric Case ($n_1=n_2$)}
The dimensionless energy transfer rate between the layers is shown in Fig.~\ref{fig1} as function of $t_2=k_B T_2/E_F$ for three different values of interaction strength ($\lambda = 1.0, 2.0, 4.0$) and three different values of $t_1=1.2, 1.0, 0.8$.  The layer separation distance is $\widetilde{d}=d/a_0=1$ for all plots.

Energy is transferred from the hot layer to the cold one, so $\mathrm{P}_{12}$ changes sign as the temperature $t_2$ crosses $t_1$. At thermal equilibrium there is no heat flow, furthermore $\mathrm{P}_{12}$ is linear in temperature difference near this point.

As we increase the interaction strength $\lambda$, the amount of transferred power in dimensionless units decreases (see Fig.~\ref{fig1}). This is due to the our definition of the scaled variables. For constant density ($k_F$), the dimensionless interaction strength $\lambda=a_0 k_F$ increases with increasing dipolar interaction. However the actual distance between the layers $d=\widetilde{d}a_0$ increases with increasing dipolar interaction as well. To sum up, although the dimensionless layer-separation distance remains constant, the actual distance $d$ varies for different values of the interaction strength $\lambda$. Note that in Fig.~\ref{fig1} as we increase the interaction strength ($\lambda = 1.0, 2.0, 4.0$) between the dipoles, we use a fixed value of ($\widetilde{d}=1$), hence the actual distance between the layers increases.

\begin{figure}
\includegraphics[width=8cm]{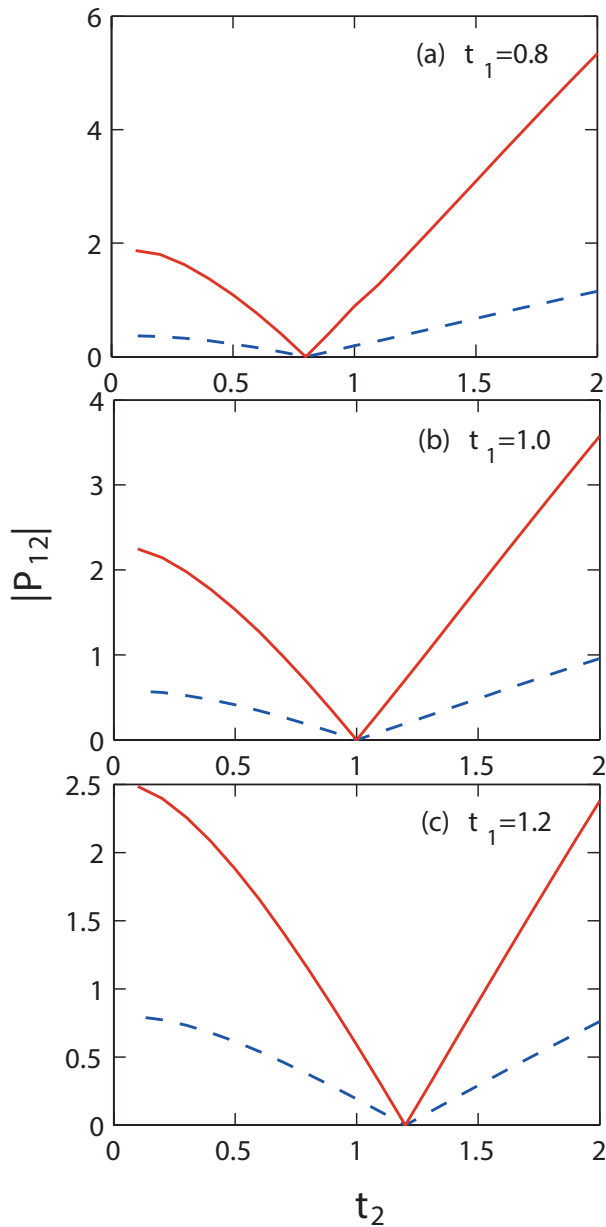}
\caption{(Color online) The interaction strength $\lambda$ dependence of the dimensionless energy transfer rate $|\mathrm{P}_{12}|$ when the actual distance between the layers is kept constant. Here, the dashed and solid lines denote the systems with $(\lambda,\widetilde{d})=(1.0,1.0)\mbox{ and }(2.0,0.5)$, respectively. The graphs are obtained for different temperature values of the first layer as $t_1 = 0.8, 1.0,1.2$, respectively.}\label{fig3}
\end{figure}

In order to isolate the effects of the dipole-dipole interaction, we investigate the system for two different values of the interaction strength $a_0$ as the actual distance $d$ is kept constant, presented in Fig.~\ref{fig3}. The systems we compare have the following parameters: $(\lambda,\widetilde{d})=(1.0,1.0)$ and $(\lambda,\widetilde{d})=(2.0,0.5)$. Here, in spite of having different scaled distances, the actual distance between the layers are the same. The energy transfer rate increases for stronger dipole dipole interactions, as expected. The increase is not quadratic in dipolar interaction ($C_{dd}$) as might be expected from a simple interpretation of Eq.~\eqref{rate}. While the rate due to the bare interaction would increase quadratically, the dynamically screened interaction and the polarization functions reduce this dependence.

\begin{figure}
\includegraphics[width=8cm]{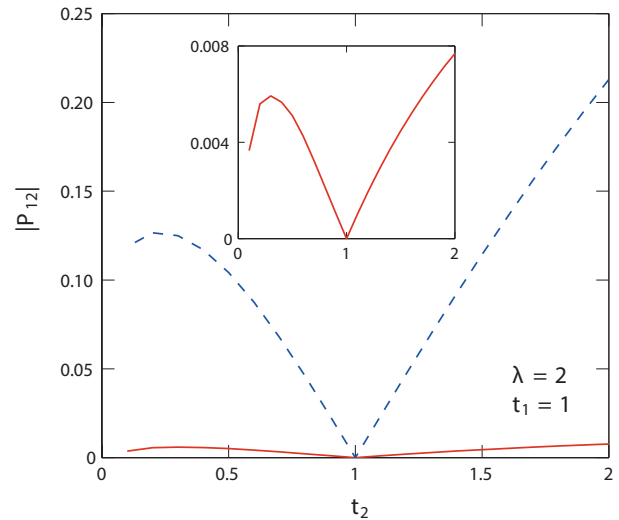}
\caption{(Color online) The well separation distance $d$ dependence of the dimensionless energy transfer rate $|\mathrm{P}_{12}|$ is presented for a constant value of interaction strength, $\lambda = 2.0$. Here, the temperature of the first layer is kept constant at Fermi value. The dashed and solid lines indicate the well separation distances for $\tilde{d}=1.0$ and $\tilde{d}=2.0$, respectively. Note that the insets show a zoomed-in view of the graphs for $\tilde{d}=2$.  }\label{fig4}
\end{figure}

We also evaluate the layer separation distance dependence of the energy transfer rate for a constant value of the interaction strength, as shown in Fig.~\ref{fig4}. Here, the first layer is at the Fermi temperature. When the distance between the layers increases, the amount of transferred energy decreases. Once again the decrease is slower than the bare interaction expectation due to screening effects.

\begin{figure}
\includegraphics[width=8cm]{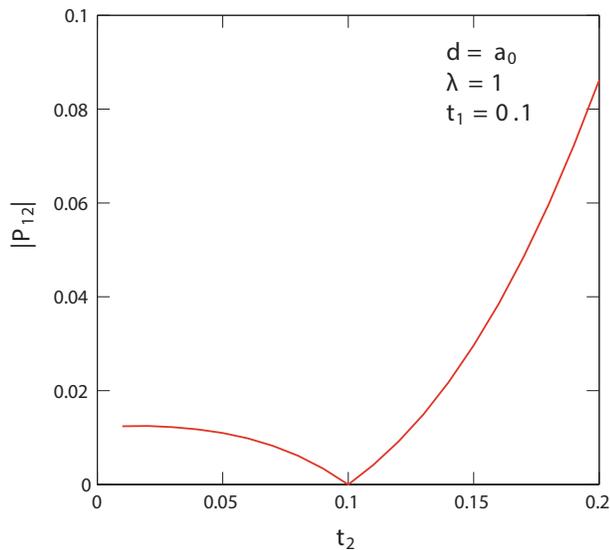}
\caption{(Color online) Low temperature limits of the system while the temperature of the first layer is kept constant at $t_1=0.1$. The interaction strength of the system is considered as $\lambda=1$ and the actual distance between the layers is equal to the length scale. Notice that the heat transfer $|\mathrm{P}_{12}|$ is an order of magnitude smaller than heat transfer obtained at $t=1$ (Fig.~\ref{fig1}). }\label{fig2}
\end{figure}

Fermion cooling gets progressively hard due to Pauli blocking as the temperature decreases. We investigated the heat transfer in our model for lower temperatures close to $0.1T_F$. The results are presented in Fig.~\ref{fig2}. Here, the interaction strength and the layer separation distance are $\lambda=1$ and $d=a_0$, respectively. As the temperature is lowered, the amount of transferred power between the layers decreases. Energy transfer between the dipoles in the opposite layers occurs due to the scattering only if there is an unoccupied final state. At low temperatures lack of unoccupied final states into which atoms can scatter suppresses heat transfer. The effect of Pauli blocking is also observable in the other figures when one of the layers is at very low temperatures.

\subsection{Asymmetric Case ($n_1\neq n_2$)}

As mentioned in Section II, the dipolar gases confined in layers with unequal densities ($n_1\neq n_2$) will have different Fermi levels. The relation equation between the densities and the interaction strengths of the layers can be calculated by $r=\sqrt{n_1/n_2}=\lambda_1/\lambda_2$, as previously defined in Eq.~\ref{r}.

\begin{figure}
\includegraphics[width=8cm]{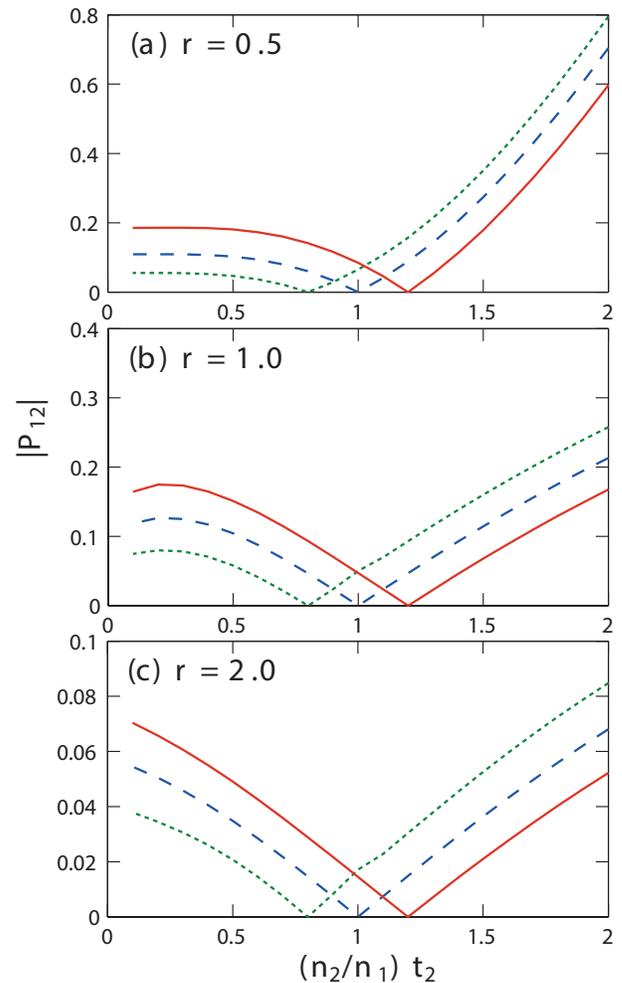}
\caption{(Color online) The dimensionless energy transfer rate $|\mathrm{P}_{12}|$ as a function of the temperature $t_2$ of the second layer for the different ratios of the interaction strengths $r=\lambda_1/\lambda_2 = 0.5, 1.0, 2.0$. Temperature of the first layer $t1$ is fixed at $t_1=1.2$ (solid lines), $t_1=1.0$ (dashed lines), and $t_1=0.8$ (dotted lines). In these graphs, the dimensionless layer separation distance of the system is considered as $\widetilde{d}=1$ and the temperature of the other layer is constant.}\label{fig5}
\end{figure}

The dimensionless energy transfer rates $|\mathrm{P}_{12}|$ as a function of the temperature $t_2$ of the second layer for different ratios of the interaction strength $r= 0.5, 1.0, 2.0$ are presented in Fig.~\ref{fig5}. The direction of the energy flow changes at thermal equilibrium points where $t_1=t_2$ for each plot. The graph (Fig.~\ref{fig5}b) obtained for $r=1$ indicates the symmetric case of the system.

Here as we decrease the density of the second layer which is the source of the heat flow in the ($t_2>t_1$) regime, the amount of transferred power decreases as expected. This effect is apparent in the graphs obtained for $r=0.5$ and $r=2$. For the low temperature values of $t_2$, once again the energy transfer is suppressed as a result of Pauli blocking.

\section{Discussion and Conclusion}
We calculate the power transferred per unit area between two parallel layers of ultracold dipolar gases which are at different temperatures. The system is characterized by two different dimensionless parameters: \textit{(i)} the interaction strength $\lambda$ and \textit{(ii)} the layer separation distance $\widetilde{d}$. In the previous section, we calculate the transferred power as a function of layer temperatures for a wide range of parameters. In this section we aim to ascertain if this contactless power transfer is an efficient cooling method.

When the two layers have the same temperature, the transferred power is zero. We can expand the transferred power $\mathcal{P}_{12}(T_1,T_2)$ around this thermal equilibrium point ($T_1=T_2$). As can be noticed in all of the plots, for a large range of temperatures $\mathcal{P}_{12}$ is well approximated by a linear fit around this point,
\begin{equation}\label{kap}
\mathcal{P}_{12} \simeq \frac{\kappa (T_2-T_1)}{d} \,.
\end{equation}
If the temperature difference between the layers is small enough for this approximation to be valid heat transfer is characterized by the slope $\kappa$ which is an effective heat conductivity. If the vacuum between the layers were filled with a material of heat conductivity $\kappa$ the transferred power per unit area would be given by Eq.\eqref{kap}. The heat conductivity will depend on the thermal equilibrium temperature around which the linear fit is carried out as well as the other parameters of the system.


We obtained the numerical values for the effective thermal conductivity for typical experimental parameters. However, the amount of heat transferred does not single-handedly determine the effectiveness of the cooling. A more transparent quantity can be obtained by a simple model of thermal dynamics between the layers.

We assume that one of the layers is kept at a constant temperature ($T_1$) by an efficient coupling to a reservoir, and investigate the temperature of the second layer as a function of time ($T_2=T_2(t)$). Energy flow from the first layer to the second one changes the internal energy of the second layer,
\begin{equation}\label{pow}
\frac{dE_2}{dt} = -\mathcal{P}_{12} \,.
\end{equation}
Relating this change to the specific heat per unit area of the Fermi gas we obtain
\begin{equation}\label{pow2}
\frac{dE_2}{dt} = C_V \frac{d}{dt}T_2(t)=C_V \frac{d}{dt} \left(T_2(t)-T_1\right)\,.
\end{equation}
If the temperature difference is small enough, the heat flow can be replaced by the linear approximation Eq.\eqref{kap}, yielding
\begin{equation}\label{pow3}
C_V \frac{d}{dt} \left(T_2(t)-T_1\right) = -\frac{\kappa (T_2(t)-T_1)}{d}\,.
\end{equation}
In this linear regime, equilibrium is approached with a time constant
\begin{equation}\label{tau}
\tau = \frac{d\, \,C_V}{\kappa}\,.
\end{equation}
For a cooling method to be effective the time constant $\tau$ must be smaller than the typical trap lifetimes.

There are two distinct classes of dipolar ultracold systems, atoms with magnetic dipoles and molecules with electric dipoles. The relevant length scales for atoms with magnetic dipoles and polar molecules are substantially different. We investigate the feasibility of using the contactless sympathetic cooling method separately for both systems.

Our calculations show that the transferred power decays rapidly with the distance between the layers. For an effective thermal coupling, the interlayer distance should be within an order of magnitude of the dipolar length scale $a_0$. For atomic species with magnetic moments $\mu$, the dipolar length scale is $a_0=\mu^2\mu_0 m/(4\pi\hbar^2)$. The length scales of three typical atoms with strong magnetic moments such as Cr, Er, Dy are calculated as $2.4$\,nm, $10.5$\,nm, $20.8$\,nm, respectively. A dipolar length scale $a_0\sim 10$\,nm means that dipolar heat transfer is effective up to at most $\sim 100$\,nm. The typical feature size of the potential in ultracold atom experiments is determined by the wavelength of the dominant transition, and is generally a few hundred nanometers. Thus the cooling scenario considered here is not directly applicable to magnetic atomic systems. It may still be possible to measure a perturbative heat transfer between very close layers.

For typical polar molecules, with electric dipole moment $p$, the dipolar length scale $a_0=p^2 m/(4\pi\varepsilon_0 \hbar^2)$ is much larger, close to a few micrometers. Thus, the creation of two layers with separation of the order of $a_0$ does not present a significant experimental difficulty.

We calculate the dipolar length scale $a_0$, the thermal conductivity $\kappa$ and the time constant $\tau$ for experimentally realized, long-lived, chemically stable Feshbach molecules, KRb~\cite{KRb}, RbCs~\cite{RbCs}, NaK~\cite{NaK1}, LiCs~\cite{LiCs1}. Our results are presented in Table~\ref{Table} for $d=a_0$ and temperatures close to the Fermi temperature for three typical densities.

The most striking result in Table~\ref{Table} is that the time constants for reaching the thermal equilibrium between the layers separated by a few micrometres is as short as tens of milliseconds. The heat transfer due to dipolar coupling is efficient for polar molecules within typical experimental distances. The cooling of ultracold polar molecules is challenging because of the extra degrees of freedom related to the rotation and vibration of the molecules. We believe the efficiency of long range heat transfer can be used to partially overcome this challenge. In particular, any cooling method can be used on only one of the layers of our model and the other layer will follow within a time scale $\tau$.

We also calculate the thermal conductivity $\kappa$ and the time constant $\tau$ at lower temperatures ($T\simeq 0.1T_F$), as shown in the Table~\ref{low}. While the thermal conductivity decreases, the specific heat of the system also decreases and time constants are not significantly affected. Dipolar thermal coupling is also effective in this low temperature regime.

\begin{widetext}
\begin{center}
\begin{table}[h]
\caption{\label{Table} Quantitative results for some ultracold polar molecules. Here, the dimensionless well separation distance is considered as $\widetilde{d}=1$. The dipole moments of the molecules are taken from references~\cite{KRb, RbCs, NaK, LiCs}. Note that $1 D = 3.34 \times 10^{-30} C\cdot m$. Here, the thermal conductivity, $\kappa$ and the time constant, $\tau$ are obtained near the thermal equilibrium with ($t_1,t_2 \simeq 1.0$)}
\begin{tabular}{c l c|c|c|c|c|}\cline{4-7}
   &                 &  & KRb  & RbCs & NaK & LiCs  \\ \hline
\multicolumn{1}{ |c  }{}  & $p$ (D)       &  & $0.57$ & $1.3$ & $2.72$ & $5.5$ \\
\multicolumn{1}{ |c  }{}   & $a_0$ ($\mu$m) &  & $0.6$  & $5.5$ &  $7.0$ & $63.4$\\
\hline
\multicolumn{1}{ |c|  } {\multirow{3}{*}{$\lambda=1$}} & \multicolumn{1}{l}{$n$ (m$^{-2}$)} &   &$2.21$ $\times 10^{11}$ & $2.63$ $\times 10^9$ & $1.62$ $\times 10^9$ & $1.98$ $\times 10^7$ \\
\multicolumn{1}{ |c|  }{}                             &  \multirow{1}{*}{$\kappa$ $\left(\hbox{W/m\ K}\right)$} &  & $1.58$  $\times 10^{-14}$  & $1.17$  $\times 10^{-17}$  & $2.0$  $\times 10^{-17}$  & $1.19$  $\times 10^{-20}$ \\
\multicolumn{1}{ |c|  }{}                             &  \multirow{1}{*}{$\tau$ (ms)} &  & $0.116$  & $17.07$   & $7.86$  & $1452$  \\\hline
\multicolumn{1}{ |c|  } {\multirow{3}{*}{$\lambda=2$}} & \multicolumn{1}{l} {$n$ (m$^{-2}$)} &   & $8.84$ $\times 10^{11}$ & $1.05$ $\times 10^{10}$ & $6.50$ $\times 10^{9}$ & $7.92$ $\times 10^{7}$ \\
\multicolumn{1}{ |c|  }{}                             &  \multirow{1}{*}{$\kappa$ $\left(\hbox{W/m\ K}\right)$} &  & $5.73$  $\times 10^{-14}$  & $4.25$  $\times 10^{-17}$  & $7.24$  $\times 10^{-17}$  & $4.33$  $\times 10^{-20}$ \\
 \multicolumn{1}{ |c|  }{}                             &  \multirow{1}{*}{$\tau$ (ms)} &  & $0.128$  & $18.81$   & $8.66$  & $1600$  \\
\hline
\multicolumn{1}{ |c|  } {\multirow{3}{*}{$\lambda=4$}} & \multicolumn{1}{l} {$n$ (m$^{-2}$)}&   & $3.54$ $\times 10^{12}$ & $4.21$ $\times 10^{10}$ & $2.60$ $\times 10^{10}$ & $3.17$ $\times 10^8$ \\
\multicolumn{1}{ |c|  }{}                             & \multirow{1}{*}{$\kappa$ $\left(\hbox{W/m\ K}\right)$} &   & $1.17$  $\times 10^{-13}$  & $8.65$  $\times 10^{-17}$  & $1.47$  $\times 10^{-16}$  & $8.81$  $\times 10^{-20}$ \\
\multicolumn{1}{ |c|  }{}                             &  \multirow{1}{*}{$\tau$ (ms)} &  & $0.251$  & $36.95$   & $17.02$  & $3144$  \\
\hline
\end{tabular}
\end{table}
\end{center}
\end{widetext}

\begin{widetext}
\begin{center}
\begin{table}[h]
\caption{\label{low} Heat transfer of some ultracold polar molecules at low temperatures ($t_1,t_2 \simeq0.1$). The quantitative results are obtained for the dimensionless systems parameters $\lambda=1$ and $\widetilde{d}=1$.}
\begin{tabular}{|l|c|c|c|c|}
\hline
                                              & $\mathrm{KRb}$  & $\mathrm{RbCs}$ & $\mathrm{NaK}$ & $\mathrm{LiCs}$  \\ \hline
$n$ (m$^{-2}$)      &$2.21$ $\times 10^{11}$ & $2.63$ $\times 10^9$ & $1.62$ $\times 10^9$ & $1.98$ $\times 10^7$ \\
$\kappa$ $\left(\hbox{W/m\ K}\right)$    & $9.61$  $\times 10^{-15}$  & $7.12$  $\times 10^{-18}$ & $1.21$  $\times 10^{-17}$ & $7.25$  $\times 10^{-21}$\\
$\tau$ (ms)                                  & $0.063$  & $9.23$ & $4.25$ & $785.5$ \\
  \hline
\end{tabular}
\end{table}
\end{center}
\end{widetext}

Sympathetic cooling method fundamentally depends on mixing two gases at different temperatures and encouraging the two to thermalize by collisions. This method is used for the cooling of neutral atoms~\cite{Myatt, Truscott, Modugno}, atomic ions~\cite{Larson} and molecular ions~\cite{Drewsen,Ostendorf}. Our calculations show that material contact between the components of the system is not necessary for sympathetic cooling, if the dipole dipole coupling is strong enough. Although we study heat transfer between two layers of fermions here, our calculations can be generalized to more complex systems. The heat transfer between the layers will be mediated by the dipolar coupling regardless of the internal dynamics of each layer. Our results provide a foundation for the future studies of cooling of ultracold dipolar gases by using heat transfer through dipolar coupling.




\acknowledgments{
We would like to thank the Scientific and Technological Research Council
of Turkey (T\"{U}B\.{I}TAK Grant no:112T974) for financial support. B.T. also thanks TUBA for support.}

\end{document}